\documentclass[10pt,twocolumn,letterpaper]{article}

\usepackage[pagenumbers]{cvpr} % To force page numbers, e.g. for an arXiv version

\usepackage{makecell}
\usepackage{multirow}
\usepackage[table]{xcolor}
\usepackage{siunitx}
\usepackage{pifont}
\usepackage{microtype}

\newcommand{\customparagraph}[1]{\smallskip\noindent\textbf{#1}\quad}
\newcommand{\cmark}{\ding{51}} % checkmark
\newcommand{\xmark}{\ding{55}} % cross

\definecolor{cvprblue}{rgb}{0.21,0.49,0.74}
\usepackage[pagebackref,breaklinks,colorlinks,allcolors=cvprblue]{hyperref}

\title{Masked Autoencoders with Limited Data: Does It Work? \\A Fine-Grained Bioacoustics Case Study}
\author{
    Wuao Liu \quad Mustafa Chasmai \quad Subhransu Maji \quad Grant Van Horn\\
    University of Massachusetts Amherst\\
    {\tt\small \{wuaoliu, mchasmai, smaji, gvanhorn\}@cs.umass.edu}\\
}

\begin{document}
\maketitle
\begin{abstract}
Bioacoustic recognition requires fine-grained acoustic understanding to distinguish similar-sounding species. 
However, many large-scale data repositories such as iNaturalist are weakly annotated, often with only a single positive species label per recording, making supervised learning particularly challenging.
Inspired by advances in computer vision, recent approaches have shifted toward self-supervised learning to capture the underlying structure of audio without relying on exhaustive annotations.
In particular, masked autoencoders (MAE) have shown strong transferability on massive audio corpora, yet their effectiveness in more modest bioacoustic settings remains underexplored.
In this work, we conduct a systematic study of MAE pretraining for species classification on iNatSounds, analyzing the impacts of pretraining data scale, domain specificity, data curation, and transfer strategies.
Consistent with prior work, we find that models pretrained on diverse general audio data achieve the best transfer performance on iNatSounds.
Contrary to observations from large-scale audio benchmarks, we find that (1) additional masked reconstruction pretraining on domain-specific data provides limited benefits and may even degrade performance relative to off-the-shelf models, and (2) selective data filtering offers a negligible advantage when the overall data scale is limited.
Our results indicate that, in moderate-sized fine-grained bioacoustic settings, pretraining scale dominates objective design. These findings further clarify when MAE-based pretraining is effective and provide practical guidance for model selection under limited supervision.
\end{abstract}   
\begin{table*}[t]
\centering
\small
\setlength{\tabcolsep}{4pt}
\begin{tabular}{lcccccccc}
\toprule
Method & SSL & SL & Pre-segmented & Pretrain Data & Domain & \#Species & Audio (h) & Eval Benchmark \\
\midrule
AudioMAE~\cite{huang2022masked} 
& \cmark & \xmark & \cmark
& AS
& General
& -
& 5.8k
& AS, ESC-50, SPC, SID \\
\midrule

BirdSet~\cite{rauch2024birdset} 
& \xmark & \cmark & \xmark 
& XC 
& Bio
& $\sim$10k
& 6.8k 
& BirdSet \\

iNatSounds~\cite{chasmai2024inaturalist} 
& \xmark & \cmark & \xmark 
& iNat 
& Bio
& 5,569
& 1.2k 
& iNat \\

Perch 2.0~\cite{van2025perch} 
& \xmark & \cmark & \cmark 
& XC, iNat, TSA, FSD50K 
& Mixed
& 14,597
& -- 
& BirdSet, BEANS \\

AVES~\cite{hagiwara2023aves} 
& \cmark & \xmark & \cmark 
& Unannotated Bio 
& Bio
& - 
& 360 
& BEANS \\

BirdMAE~\cite{rauch2025maskedautoencoderslistenbirds} 
& \cmark & \xmark & \cmark 
& XC 
& Bio
& $\sim$10k
& 6.8k 
& BirdSet \\

\midrule

AVEX~\cite{miron2025matters} 
& \cmark & \cmark & \cmark 
& XC, iNat, AS 
& Mixed
& $\sim$15k 
& -- 
& BirdSet, BEANS  \\

\rowcolor{gray!10}
\textbf{Ours} 
& \cmark & \cmark & \xmark 
& iNat, AS
& Mixed
& 5,569
& 1.2k 
& iNat \\
\bottomrule
\end{tabular}
\caption{
    \textbf{Comparison of training data across prior bioacoustic models.}
    Our model is pretrained on both general-domain and bioacoustic audio, and supports audio recordings of arbitrary length.
    Abbreviations: AS = AudioSet; SID = speaker identification on VoxCeleb; XC = Xeno-Canto; iNat = iNatSounds; TSA = Tierstimmenarchiv; SL = supervised learning; SSL = self-supervised learning.
}
% ESC-50: Environmental Sound Classification; SPC: Speech Commands;
\label{tab:data_stats}
\end{table*}
\section{Introduction}
\label{sec:intro}
Monitoring biodiversity at scale is critical for understanding ecosystem health~\cite{diaz2019pervasive}, tracking species decline~\cite{butchart2010global}, and informing conservation efforts~\cite{schmeller2015towards}.
Passive acoustic monitoring (PAM) systems enable continuous, non-invasive recording of environmental soundscapes, offering an efficient way to survey wildlife across vast and remote regions~\cite{sugai2019terrestrial, gibb2019emerging, ross2023passive}. 
However, translating these recordings into reliable species-level information requires models capable of distinguishing subtle, fine-grained acoustic differences under highly variable and noisy conditions~\cite{chasmai2024inaturalist}.
Bioacoustic classification has therefore attracted significant attention in recent years~\cite{chasmai2024inaturalist, van2025perch, rauch2025maskedautoencoderslistenbirds, hagiwara2023aves}.

\begin{figure}[t]
    \centering
    \includegraphics[width=\linewidth]{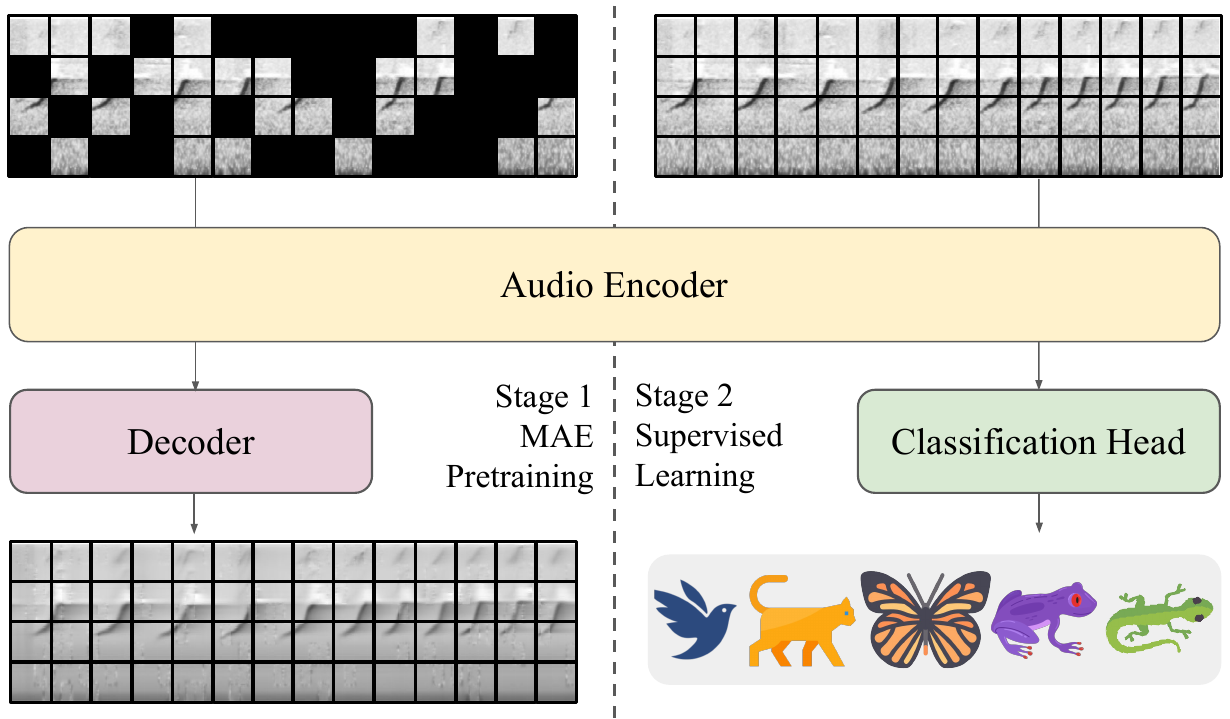}
    \caption{\textbf{Overview.} We investigate MAE pretraining for fine-grained bioacoustic recognition. An audio encoder is first trained with a masked spectrogram reconstruction objective and then finetuned for species classification. We systematically evaluate its effectiveness under a modest data regime and find that pretraining scale plays a more critical role than continual pretraining.}
    \label{fig:overview}
\end{figure}

A common approach in bioacoustics is to apply vision architectures to time–frequency representations of audio, such as spectrograms~\cite{chasmai2024inaturalist, sunmerlin, van2022exploring}. In practice, bioacoustic datasets are typically smaller and noisier than their image counterparts~\cite{van2018inaturalist, maji2013fine, KrauseStarkDengFei-Fei_3DRR2013, fu2017look, lin2015bilinear}, further amplifying the challenges of fine-grained recognition.
Moreover, acquiring large-scale bioacoustic data remains difficult, as annotation requires substantial domain expertise and extensive manual effort. The inherent complexity of natural soundscapes---with overlapping vocalizations and low signal-to-noise ratios---poses a fundamental scalability bottleneck for supervised learning.

Recent efforts have instead turned to citizen science platforms like Xeno-Canto~\cite{xeno} and iNaturalist~\cite{iNaturalist} to obtain broader species and geographical coverage.
Examples include BirdSet~\cite{rauch2024birdset}, which curates bird vocalizations from Xeno-Canto, and iNatSounds~\cite{chasmai2024inaturalist}, which contains recordings of a wide range of species worldwide.
The downside, however, is that such datasets primarily consist of curated focal recordings with positive-only annotations, leading to domain gaps and increased label noise. 

Self-supervised learning (SSL) has emerged as a dominant paradigm for representation learning, particularly when labeled data is limited.
Among recent SSL approaches, masked autoencoders (MAEs) have achieved remarkable success in image~\cite{he2022masked, wei2023diffusion}, video~\cite{feichtenhofer2022masked, tong2022videomae, wang2023videomae, naiman2025lv}, and general audio tasks~\cite{huang2022masked, baade2022mae}.
By training models to reconstruct held-out patches in an image (or audio spectrogram), MAEs learn representations that capture the underlying structure of the input, enabling effective transfer to downstream tasks. 
Applications of MAEs span remote sensing~\cite{reed2023scale, cong2022satmae, tang2023cross}, medical image processing~\cite{Wald_2025_CVPR}, and autonomous driving~\cite{Lin_Wang_Qi_Dong_Yang_2024}, among others. This success motivates the central question of our work: Can MAEs be effectively adapted for representation learning in bioacoustics?

In this paper, we explore pretraining MAEs on moderate-sized bioacoustic data and evaluate their transferability to downstream species classification tasks. 
SSL has seen some success in bioacoustics~\cite{miron2025matters}, including earlier exploration of MAEs~\cite{hagiwara2023aves, rauch2025maskedautoencoderslistenbirds}. We reconfirm some of their findings, such as the benefit of including diverse audio recordings during pretraining. Our results contrast some other findings, however, and highlight potential limitations in moderate-sized data settings such as the iNatSounds dataset (5x lesser audio, 3x fewer species than prior work; Table~\ref{tab:data_stats}). We find that additional MAE pretraining on domain-specific data provides limited gains and sometimes degrades performance relative to off-the-shelf models. Selective data curation also offers a negligible advantage when the overall data scale is limited.
Our experiments probe the effects of various pretraining and post-training strategies and provide key insights into when these approaches are effective.
\section{Related Work}
\label{sec:related}
\subsection{Self-Supervised Learning for Audio}
Existing SSL methods in audio are typically categorized into contrastive learning (CL) and masked spectrogram modeling (MSM) approaches. Contrastive learning structures the embedding space such that semantically related audio instances are mapped nearby, while unrelated signals are separated. Early works such as wav2vec~\cite{schneider2019wav2vec,baevski2020wav2vec} and HuBERT~\cite{hsu2021hubert} demonstrated that learning from large-scale unlabeled speech via contrastive or clustering-based objectives yields strong representations for speech recognition and general audio tasks.
Several bioacoustic encoders follow this paradigm, including Aves~\cite{hagiwara2023aves} and its extensions EAT~\cite{chen2024eat} and BEATs~\cite{chen2023beats}.
However, in fine-grained bioacoustic settings, constructing semantically meaningful positive pairs is less straightforward. Recordings of the same species can vary substantially across individuals, geographic regions, and behavior contexts~\cite{chasmai2024inaturalist}, introducing significant intra-class variability that may weaken the invariance assumptions underlying contrastive objectives.
Recent work further extends contrastive learning to cross-modal settings inspired by CLIP~\cite{radford2021learning}. Models such as CLAP~\cite{elizalde2023clap} and AudioCLIP~\cite{guzhov2022audioclip} align audio with text for improved transferability, while BioLingual~\cite{robinson2024transferable} leverages captioned wildlife recordings to enable zero-shot classification and retrieval.

In parallel, masked or generative modeling approaches rely on reconstructing partially observed audio inputs as pretext tasks. These methods either reconstruct masked portions of the input signals or recover masked features in a latent space. Inspired by masked image modeling, MAEs have been adapted to audio domain via masked spectrogram modeling. AudioMAE~\cite{huang2022masked} demonstrates that reconstructing masked spectrogram patches can yield competitive representations on AudioSet~\cite{audioset}, outperforming supervised pretraining baselines by up to 6.0 mAP on AudioSet-2M. 

Recent work in audio foundation models expands beyond classification to broader perception tasks. For example, SAM-Audio~\cite{shi2025sam} is a unified audio separation model that learns to isolate arbitrary sounds from complex mixtures using text, visual, or temporal span prompts, offering flexible sound extraction across speech, music, and general sounds. 

Despite these advances, most evaluations of masked audio models focus on general audio event classification or trimmed audio clips. Their effectiveness for fine-grained species recognition and in-the-wild audio recordings under long-tailed and weakly annotated bioacoustic conditions remains less explored. Our work provides a systematic examination of MAE pretraining in this challenging setting.

\subsection{Bioacoustics}
The field of bioacoustics has seen significant interest and progress in recent years. Large-scale datasets such as BirdSet~\cite{rauch2024birdset} have spurred research, enabling the development of strong foundation models~\cite{van2025perch, miron2025matters, schwinger2025foundation}. 
\citet{van2025perch} demonstrated the potential of supervised learning with a mixed training dataset consisting of diverse taxa as well as general audio from FSD50K~\cite{fonseca2021fsd50k}. BirdNet~\cite{kahl2021birdnet} is another bioacoustics model boasting broad applicability and strong generalization to downstream domains. Previous work has also developed datasets and models targeting specific taxonomic groups, including birds~\cite{kahl2021birdnet, kryklyvets2025mavis}, amphibians~\cite{genevclassification, moummad2024mixture}, insects~\cite{faiss2025insectset459}, as well as marine mammals~\cite{bae2025entropy}.

BirdMAE~\cite{rauch2025maskedautoencoderslistenbirds} was the first to explore the use of MAEs in bioacoustics, and achieved strong results when pretrained on Xeno-Canto data. 
While it focuses more on the pretraining protocol and objective design with a 1.2M bioacoustic dataset, we highlight efficient learning when the goal is to transfer a large foundation models to smaller target datasets and study the recipes to optimize the models. This setting better reflects practical deployment scenarios in ecological research.
More recently, \citet{miron2025matters} has also shown promising results with SSL followed by standard supervised fine-tuning on a large-scale dataset consisting of Xeno-Canto, iNaturalist, and AudioSet~\cite{audioset}.
Our study differs in two key aspects. First, we explore SSL on iNatSounds alone, which is considerably smaller than prior large-scale corpora. Second, while these approaches focus solely on out-of-domain evaluations, we evaluate in-domain performance with the iNatSounds test set. Our results shed light on what may be needed for SSL to work, and cases where it actually benefits performance.

Beyond species classification, bioacoustic research has investigated related tasks including individual identification~\cite{miron2025matters}, call-type recognition~\cite{miron2025matters}, life-stage detection~\cite{robinson2025naturelmaudio}, and geo-localization~\cite{chasmai2025audio}. 
NatureLM-Audio~\cite{robinson2025naturelmaudio} extends this line of work by enabling free-form responses to bioacoustic queries and represents the first large-scale audio-language model tailored to bioacoustics. 
The category-agnostic representations learned by MAEs may further benefit these downstream tasks.

\section{Problem Setup and Methodology}
\label{sec:prelim}

\paragraph{Problem Setting.}
Let $\mathbf{x} \in \mathcal{X}$ denote an audio recording and $\mathbf{y} \in \{1, \cdots, C\}$ denote its species label, where $C$ is the number of species. Although bioacoustic classification is inherently a multi-label problem due to overlapping vocalizations, the available annotations only specify a single primary species per recording. We therefore formulate the task as a single-label classification problem. The objective is to learn a model $g : \mathcal{X} \rightarrow \{1, \cdots, C\}$ for fine-grained bioacoustic classification.
We adopt a two-stage training paradigm: (i) self-supervised pretraining on large-scale unlabeled audio data, and (ii) adaptation to a smaller labeled target dataset. We decompose the model as
\[
g = h_{\boldsymbol{\phi}} \circ f_{\boldsymbol{\theta}},
\]
where $f_{\boldsymbol{\theta}} : \mathcal{X} \rightarrow \mathcal{Z}$ is an audio encoder that maps an audio recording to a latent representation $\mathbf{z} \in \mathcal{Z}$, and $h_{\boldsymbol{\phi}} : \mathcal{Z} \rightarrow \{1, \cdots, C\}$ is a classification head.
The encoder $f_{\boldsymbol{\theta}}$ is first pretrained on an unlabeled dataset
\[
\mathcal{D} = \{\mathbf{x}_i\}_{i=1}^{N},
\]
with a self-supervised masked reconstruction objective. The pretrained encoder is then adapted to a domain-specific labeled dataset of a smaller size 
\[
\hat{\mathcal{D}} = \{(\hat{\mathbf{x}}_j, \hat{\mathbf{y}}_j)\}_{j=1}^{M}, M \ll N.
\]

During downstream adaptation, we consider the following strategies:
\begin{itemize}
    \item \textbf{Linear probing:} The encoder parameters $\boldsymbol{\theta}$ are frozen, and only the classifier $h_{\boldsymbol{\phi}}$ is trained.
    \item \textbf{Finetuning:} Both $\boldsymbol{\theta}$ and $\boldsymbol{\phi}$ are updated using cross-entropy loss on $\hat{\mathcal{D}}$.
\end{itemize}

\begin{figure*}[t]
    \centering
    \includegraphics[width=\linewidth]{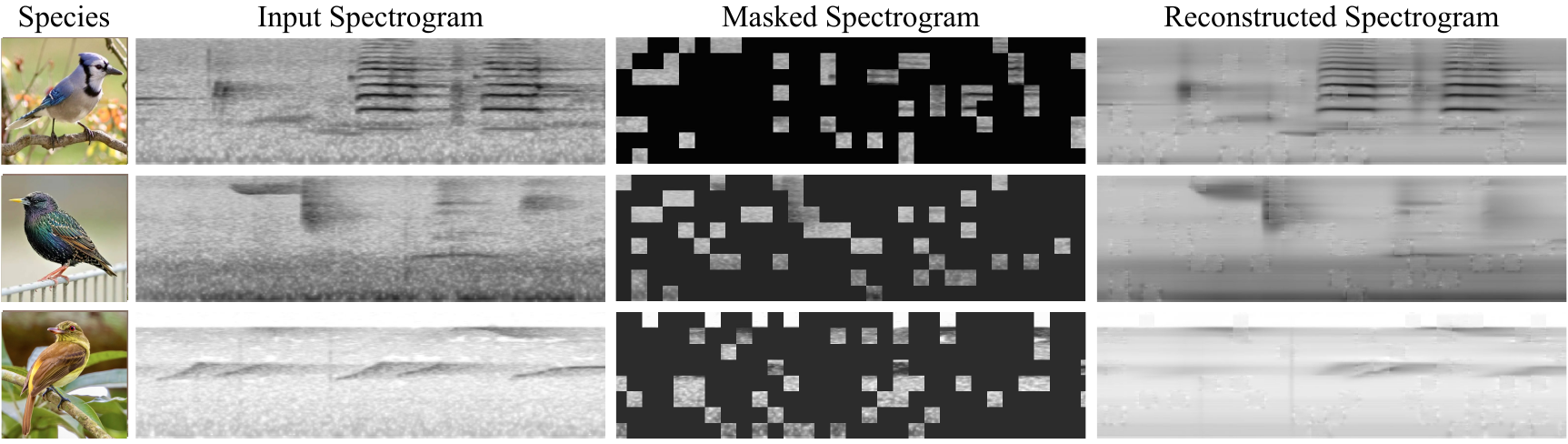}
    \caption{
    \textbf{Reconstruction performance.}
    Qualitative examples of masked spectrogram reconstruction for three species in the iNatSounds validation set.
    We use a ViT-B encoder following AudioMAE's default configuration.
    Input spectrograms are masked with a ratio of 0.8.
    The model directly predicts masked pixel values for visualization.
    }
    \label{fig:visualisations}
\end{figure*}

\paragraph{Masked Autoencoders for Audio.}
Given an audio recording $\mathbf{x}$, we first convert it into a time--frequency representation (log-mel spectrogram). The spectrogram is treated as a 2D image and partitioned into non-overlapping patches. Let $\mathbf{X} \in \mathbb{R}^{H \times W}$ denote the spectrogram and $\{\mathbf{p}_k\}_{k=1}^{K}$ denote the resulting patches. A random masking operator selects a subset of $\mathcal{M}$ patches to mask and only the visible patches $\{\mathbf{p}_k : k \notin \mathcal{M}\}$ are fed into the audio encoder $f_{\boldsymbol{\theta}}$, producing latent representations for the visible subset.
A lightweight decoder $d_{\boldsymbol{\psi}}$ then inserts learnable mask tokens at masked positions, restores the original patch ordering, and predicts pixel-level reconstructions of all patches.

The pretraining objective minimizes the reconstruction loss over the masked patches:
\[
\mathcal{L}_{\text{MAE}} = 
\frac{1}{|\mathcal{M}|}
\sum_{k \in \mathcal{M}}
\left\| \mathbf{p}_k - \hat{\mathbf{p}}_k \right\|_2^2.
\]

Following AudioMAE, we adopt a patch-wise normalized reconstruction target rather than raw spectrogram values. Specifically, each patch is normalized independently before computing the mean squared error (MSE) loss. This normalization stabilizes training and has been empirically shown to improve the quality of learned representations for downstream classification tasks. After pretraining, the decoder $d_{\boldsymbol{\psi}}$ is discarded, and only the encoder $f_{\boldsymbol{\theta}}$ is retained for transfer learning.
\section{Experiments}
\label{sec:exps}
In this section, we describe our experimental setup (Section~\ref{sec:data}-\ref{sec: impl}) and report our results. Our experiments explore MAE pretraining paradigm from three lenses: 1) post-training strategy (Sections~\ref{sec:probe}-\ref{sec:partial_finetune}), 2) pretraining domain (Section~\ref{sec:pretrain_domain}), and 3) pretraining data quality (Section~\ref{sec:curation}). 

\subsection{Datasets and Evaluation}
\label{sec:data}
\customparagraph{Datasets.} We conduct our experiments on both iNatSounds~\cite{chasmai2024inaturalist} and AudioSet~\cite{audioset} datasets.
AudioSet contains approximately 2 million 10-second audio clips collected from YouTube, each annotated with one or more labels from 527 audio event classes. Although labels are not used during pretraining, we follow the official split and download the Hugging Face version, filtering out corrupted recordings. This results in 18,373 balanced training clips, 1,861,646 unbalanced training clips, and 17,148 evaluation clips. We combine the balanced and unbalanced sets for pretraining.
iNatSounds is a weakly annotated bioacoustic dataset sourced from iNaturalist~\cite{iNaturalist}, a global citizen science platform. It contains approximately 137K, 45K, and 49K audio recordings in the training, validation, and test splits, respectively. The training split covers 5,569 species, while the validation and test splits include a subset of 1,212 species.

\customparagraph{Evaluation Protocol.}
We sample audio recordings from the training splits of AudioSet and iNatSounds for pretraining, and use only the labeled audio recordings from the iNatSounds training split for finetuning.
Evaluation is conducted at the ``file-level''. Regardless of its duration, each recording produces a single prediction over the 1,212 evaluation species. We report class-averaged Top-1 and Top-5 accuracy on the iNatSounds validation and test splits.

\subsection{Implementation Details}
\label{sec: impl}

\begin{table}[b]
    \centering
    \caption{
    \small
    \textbf{Linear probing results on iNatSounds.} We report Top-1 and Top-5 classification accuracy (\%) on validation and test sets using frozen pretrained encoders.
    }
    \label{tab:linear_mae}
    \begin{tabular}{ccccc}
    \toprule
    \multirow{2}{*}{Method} & \multicolumn{2}{c}{iNat Val} & \multicolumn{2}{c}{iNat Test}\\
    & Top-1 & Top-5 & Top-1 & Top-5\\
    \midrule
    Chance   & 0.08 & 0.41 & 0.08 & 0.41 \\
    BirdMAE  & 2.48 & 5.42 & 2.40 & 5.10 \\
    AudioMAE & 3.63 & 8.16 & 3.27 & 7.62 \\
    Ours     & 3.65 & 8.53 & 3.54 & 8.06 \\
    \bottomrule
\end{tabular}
\end{table}

\begin{table*}[t]
    \centering
    \caption{
    \textbf{Finetuning results on iNatSounds and ablation study on the pretraining data.}
    Left: MAE models pretrained with a different combination of audio datasets, including AudioSet and iNatSounds, and finetuned on iNatSounds species classification task. We denote the MAE model initialized with random weights and continual pretrained on iNatSounds as iNat-SSL, shown in row 3.
    Right: We show the input, the masked and the reconstructed spectrograms of MAE variants. Starting from the second row, the reconstruction results are from our full MAE, a MAE trained with only iNatSounds data, and a MAE trained when the decoder weights are frozen. The best reconstruction is obtained when the model is trained with both general audio data and fine-grained bioacoustic data, and has a trainable decoder.
    SL: supervised learning. SSL: self-supervised learning, pretrained using MAE pipeline. IN: ImageNet. AS: AudioSet. iNat: iNatSounds.
    }
    \label{tab:main_ft}
    \begin{minipage}[t]{0.55\textwidth}
        \setlength{\tabcolsep}{4pt}
        \renewcommand{\arraystretch}{1.12}
        \begin{tabular}{cccccc}
            \toprule
            \multirow{2}{*}{Initialization} & \multirow{2}{*}{Continue Pretrain} & \multicolumn{2}{c}{iNat Val} & \multicolumn{2}{c}{iNat Test}\\
            & & Top-1 & Top-5 & Top-1 & Top-5 \\
            \midrule
            IN-SL  & \xmark & 58.76 & 77.62 & 57.69 & 77.51 \\
            \midrule
            Random & \xmark & 38.02 & 58.19 & 37.13 & 57.01 \\
            Random & \cmark & 49.03 & 68.67 & 47.53 & 68.11 \\
            $\text{AS-SSL}^{rel}$ & \xmark & 58.65 & 78.53 & 57.54 & 78.06 \\
            $\text{AS-SSL}^{rel}$ & \cmark & 59.51 & 78.63 & 58.10 & 78.65 \\
            \midrule
            BirdMAE & \xmark & 63.61 & 82.79 & 62.76 & 82.58 \\
            \bottomrule
        \end{tabular}
    \end{minipage}
    \begin{minipage}[c]{0.43\textwidth}
    \vspace{0pt}
    \includegraphics[
      width=\linewidth,
    ]{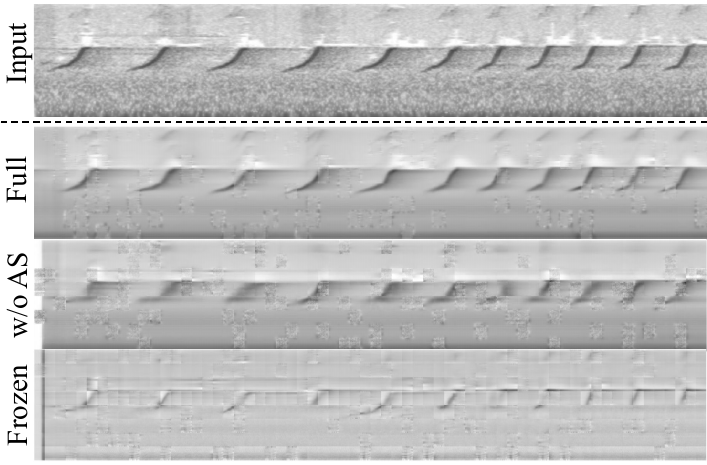}
    \end{minipage}
\end{table*}

\customparagraph{Spectrogram Generation.} Following prior work~\cite{chasmai2024inaturalist, huang2022masked}, we convert raw audio into log-mel spectrograms using STFT~\cite{pedersen1965mel} with a window size of 512 and hop size of 128. Frequencies from 50 Hz to 11.025 kHz are mapped to 128 mel bins. Each recording is segmented into overlapping 3-second clips with a stride of 1.5 seconds, resulting in spectrograms of size $128 \times 512 \times 1$. Please see Fig~\ref{fig:visualisations} for a few exemplar spectrograms. 

\customparagraph{Model Configuration.} We adopt the AudioMAE~\cite{huang2022masked} architecture, using a 12-layer ViT-B encoder and a 16-layer transformer decoder with shifted local attention. The model is first pretrained with the masked reconstruction objective and then transferred to the downstream species classification task. We use a base learning rate of $1 \times 10^{-3}$, weight decay of $1 \times 10^{-4}$, and train the models for 100 epochs during pretraining stage. All experiments are conducted on a single NVIDIA A100 GPU. Pretraining takes approximately 24 hours while fine-tuning requires around 16 hours.

\subsection{MAEs Underperform Under Linear Probing}
\label{sec:probe}
We first evaluate the quality of pretrained representations using a standard linear probing protocol.
In Tab.~\ref{tab:linear_mae}, MAE models yield extremely low classification performance across all variants. AudioMAE achieves 3.27\% Top-1 accuracy on the test set, while BirdMAE further drops to 2.40\%. Our variant slightly improves upon AudioMAE, reaching 3.54\% Top-1 and 8.06\% Top-5 accuracy on the test split, but the overall performance remains low. Notably, these results remain poor even when pretraining is conducted on iNatSounds itself and evaluated on the same dataset via frozen features.

These findings suggest that the masked reconstruction objective alone does not produce linearly separable features suitable for fine-grained bioacoustic recognition. Features learned from general audio training may fail to capture the fine-grained details essential for species classification, and the markedly low performance underscores the inherent difficulty of the problem. Unlike observations reported on visual pretraining~\cite{he2022masked}, MAE-pretrained representations on iNatSounds appear to require substantial task-specific adaptation, motivating further investigation into fine-tuning strategies.

\subsection{Finetuning Emphasizes Pretraining Scale}
\label{sec:finetune}
We next evaluate full finetuning on iNatSounds. A randomly initialized model achieves 37.13\% Top-1 and 57.01\% Top-5 accuracy on the test split (Table~\ref{tab:main_ft}, second row), establishing a lower bound. Pretraining on iNatSounds with masked spectrogram reconstruction improves Top-1 accuracy by 10.4 percentage points, demonstrating that masked spectrogram modeling provides a useful initialization even at a moderate scale.
However, models pretrained on larger and more diverse datasets substantially outperform iNat-SSL continual pretraining. Both ImageNet-supervised (IN-SL) and AudioSet self-supervised (AS-SSL) pretraining improve performance by approximately 10 percentage points in both Top-1 and Top-5 accuracy over iNat-SSL. This indicates that iNatSounds alone is insufficient for learning strong representations from scratch, and pretraining scale and diversity play a more critical role than strict domain alignment.

In contrast to prior findings~\cite{huang2022masked}, we observe that AudioSet-pretrained MAE models do not consistently outperform ImageNet-supervised MAE models on iNatSounds. 
With a roughly 0.1 percentage point gap in Top-1 accuracy between AS-SSL and IN-SL, performance appears to be driven more by pretraining scale than by data modality in this moderate-scale bioacoustic setting.

The best performance is achieved by first pretraining on AudioSet and then further adapting with masked reconstruction on iNatSounds before a full finetuning. However, the additional gain from domain-specific self-supervised adaptation is small, reinforcing that most improvements stem from large-scale general pretraining rather than in-domain masked reconstruction.

Finally, we finetune BirdMAE, which is pretrained on substantially larger and more diverse bioacoustic corpora. It achieves the strongest performance among all MAE variants (63.61\% and 62.76\% Top-1 accuracy under two settings), further supporting the conclusion that scaling data volume and diversity is the dominant factor for downstream transfer. Notably, BirdMAE is trained with approximately 6$\times$ longer audio duration and 2$\times$ more species than in our setting, further emphasizing that data scale and diversity are critical for effective bioacoustic pretraining.

\begin{figure}[t]
    \centering
    \includegraphics[width=0.98\linewidth]{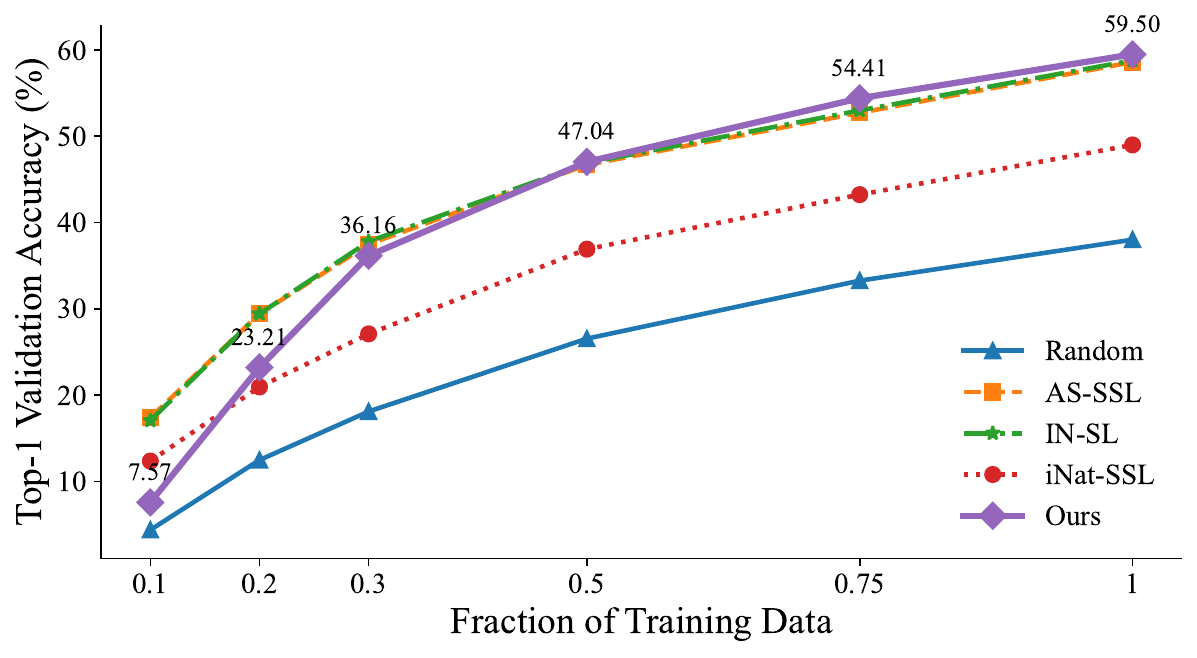}
    \caption{\textbf{Finetuning MAEs using different fractions of bioacoustic data}. We show Top-1 validation accuracy on iNatSounds as a function of the available labeled samples.}
    \label{fig:partial_finetune}
\end{figure}

\subsection{Partial Finetuning and Sample Efficiency}
\label{sec:partial_finetune}
While full finetuning measures final downstream task performance, it does not assess how efficiently pretrained representations utilize labeled data. Since fine-grained bioacoustic annotations are limited and costly, we evaluate the \emph{sample efficiency} of different pretraining strategies by finetuning on varying fractions of the iNatSounds training set. We randomly subsample the training data at different ratios and finetune all MAEs under the same optimization settings. 

Fig.~\ref{fig:partial_finetune} shows that pretrained models substantially outperform training from random initialization across all data fractions, confirming the importance of MAE pretraining.
However, general-audio MAE pretraining (AS-SSL) does not consistently outperform the ImageNet-supervised baseline (IN-SL). Despite being pretrained on natural images rather than audio data, IN-SL exhibits competitive and in several regimes slightly stronger sample efficiency. Pretraining on iNatSounds alone yields moderate gains over random initialization but remains inferior to large-scale general audio or ImageNet pretraining. Combining general audio and domain-specific pretraining (AS-SSL + iNat-SSL) provides the strongest overall performance, though the improvements over single-stage pretraining are still modest. Overall, these results suggest that, at the scale of iNatSounds, MAE pretraining on general audio does not provide clear advantages in adaptation efficiency over strong supervised vision pretraining. The scaling trends indicate that initialization quality, rather than domain alignment alone, plays a dominant role in downstream performance under limited supervision.

\subsection{Pretraining on Mixed General and Bioacoustic Audio Recordings}
\label{sec:pretrain_domain}
According to~\cite{huang2022masked}, audio transformer models benefit significantly from diverse training data. To further explore this finding in the context of MAEs, we examine if adding general-purpose audio with fine-grained bioacoustic recordings improves species classification on iNatSounds. Thus, we train MAEs from scratch using various mixtures of AudioSet and iNatSounds training samples.

\begin{table}[ht]
    \centering
    \setlength{\tabcolsep}{10pt}
    \caption{\textbf{Pretraining on mixed AudioSet and iNatSounds data.} 
    We fix the total batch size during MAE pretraining and vary the sampling ratio between AudioSet and iNatSounds. All models are then fine-tuned on iNatSounds and evaluated on the validation split using class-averaged Top-1 and Top-5 accuracy.
    }
    \begin{tabular}{lcc}
    \toprule
    Data Composition & \multicolumn{2}{c}{iNat Val}\\
    (AudioSet: iNat) & Top-1 & Top-5 \\
    \midrule
    iNat-only  & 49.0 & 68.7 \\
    \midrule
    \(1{:}1\)   & 53.8 & 74.7 \\
    \(3{:}1\)   & 57.5 & 77.4 \\
    \(7{:}1\)   & 59.4 & 78.9 \\
    \(15{:}1\)  & 60.1 & 80.3 \\
    \midrule
    AS-only    & 60.2 & 80.2 \\
    \bottomrule
    \end{tabular}
    \label{tab:co_training}
\end{table} 

As shown in Table~\ref{tab:co_training}, we observe a surprising result that pretraining with general-purpose audio from AudioSet outperforms pretraining on the iNatSounds dataset alone. Validation accuracy consistently improves as the proportion of AudioSet recordings sampled per batch increases. Specifically, pretraining on iNatSounds alone yields a Top-1 accuracy of 49.0\%, while using only AudioSet achieves 60.2\%, representing a nearly 10 percentage points improvement. Since the total number of training samples is held constant, it suggests that general audio is a more effective pretraining source for the masked modeling objective. One possible explanation is that general audio recordings tend to exhibit more complex patterns and richer textures in their spectrograms, whereas bioacoustic recordings often contain more background noise or long periods of silence.

\begin{figure}[t]
    \centering
    \includegraphics[width=\linewidth]{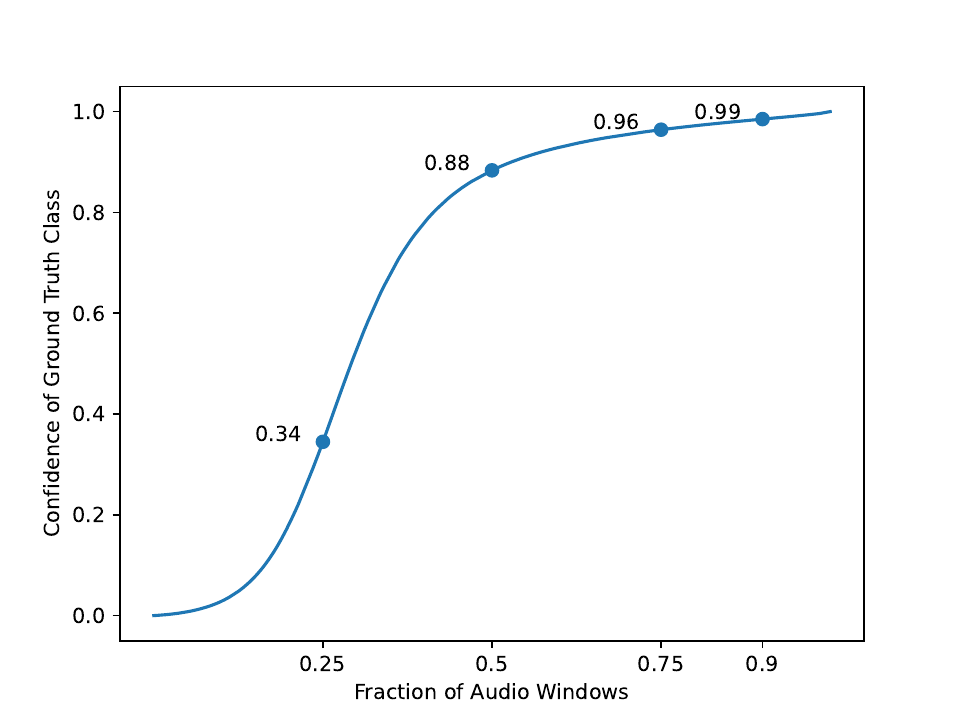}
    \begin{minipage}{0.58\linewidth}
        \small
        \setlength{\tabcolsep}{1pt}
        \begin{tabular}{c c c c}
        \toprule
        \makecell{Conf. \\ Thres.} & \makecell{Segments \\ Dropped (\%)} & \makecell{Avg Drop \\ Per File (\%)} & \makecell{MV3 \\ Acc.} \\
        \midrule
        0.00 & 0  & 0  & 54.9 \\
        0.30 & 25 & 18 & 53.7 \\
        0.50 & 30 & 30 & 53.2 \\
        0.90 & 50 & 45 & 54.1 \\
        0.95 & 75 & 60 & 53.4 \\
        \bottomrule
        \end{tabular}
    \end{minipage}
    \begin{minipage}[t]{0.4\linewidth}
        \raggedleft
        \small
        \setlength{\tabcolsep}{1pt}
        \begin{tabular}{c c c}
        \toprule
        \makecell{Conf. \\ Thres.} & \makecell{Filtered \\ Ft Acc.} & \makecell{Full \\ Ft Acc.} \\
        \midrule
        0.00 & - & 59.5 \\
        0.50 & 56.1 & 54.9 \\
        0.90 & 54.9 & 54.8 \\
        \bottomrule
        \end{tabular}
    \end{minipage}
    \hfill
    \caption{
    \textbf{Audio segments filtered by classification confidence.} Left: Distribution of confidence scores across audio segments. Middle: Fraction of audio segments to keep under different confidence thresholds, along with the corresponding validation accuracy using a MobileNetV3 model trained on the filtered data. Right: Top-1 validation accuracy of ViT-B models fine-tuned on both iNatSounds full dataset and filtered subsets.
    }
    \label{fig:classification_scores}
\end{figure}

\begin{figure}[t]
    \includegraphics[width=\linewidth]{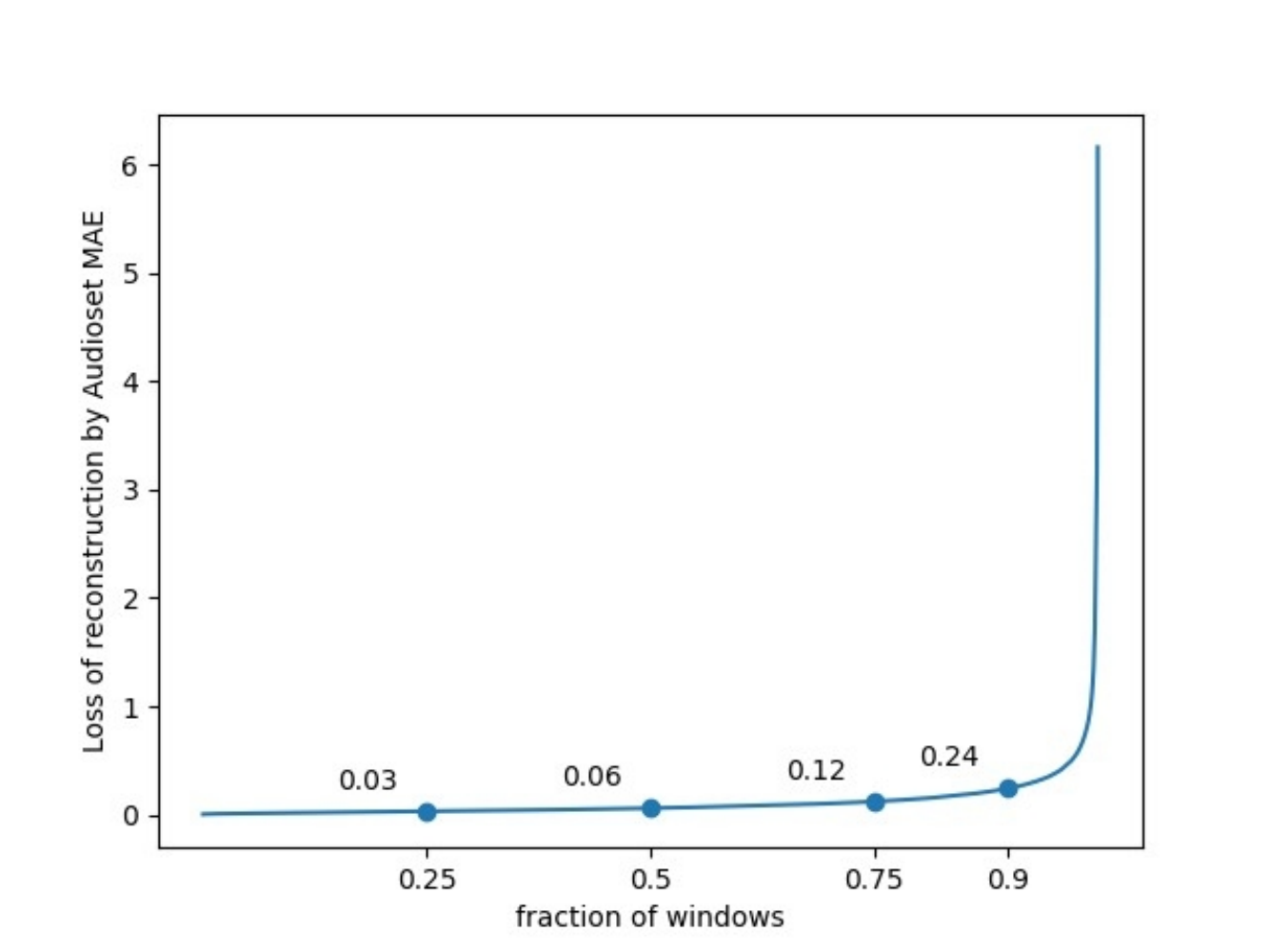}
    \begin{minipage}{0.49\linewidth}
        \centering
        \small
        \setlength{\tabcolsep}{3pt}
        \begin{tabular}{c c c}
        \toprule
        \makecell{Conf. \\ Thres.} & \makecell{Segments \\ Dropped (\%)} & \makecell{MV3 \\ Acc.} \\
        \midrule
        0.00 & 0 & 54.9 \\
        0.06 & 50 & 50.5 \\
        0.12 & 75 & 45.9 \\
        0.25 & 90 & 44.9 \\
        \bottomrule
        \end{tabular}
    \end{minipage}
    \begin{minipage}{0.49\linewidth}
        \raggedleft
        \small
        \setlength{\tabcolsep}{3pt}
        \begin{tabular}{c c c}
        \toprule
        \makecell{Conf. \\ Thres.} & \makecell{Segments \\ Dropped (\%)} & \makecell{Full \\ Ft Acc.} \\
        \midrule
        0.00 & 0 & 59.5 \\
        0.06 & 25 & 57.9 \\
        0.12 & 50 & 55.7 \\
        0.25 & 75 & 52.2 \\
        \bottomrule
        \end{tabular}
    \end{minipage}
    \caption{
    \textbf{Audio segments filtered by reconstruction loss.} Left: Distribution of confidence scores across audio segments. Middle: Fraction of audio segments to keep under different confidence thresholds, along with the corresponding validation accuracy using a MobileNetV3 model trained on the filtered data. Right: Top-1 validation accuracy of ViT-B models fine-tuned on both iNatSounds full dataset and filtered subsets.
    }
    \label{fig:reconstruction_loss}
\end{figure}

\subsection{Pretraining on Curated Recordings}
\label{sec:curation}

A potential hypothesis for the modest gains of MAE pretraining is the quality of the pretraining data. iNatSounds suffers from its weak labels, where species annotations are provided at the recording level, and the typical sliding-window approach yields a large number of empty audio segments. This leads to potential mislabeling for supervised learning approaches, and may also negatively affect pretraining, as a substantial portion of the training objective is devoted to reconstructing background noise rather than informative acoustic events. The impact of such noisy data is likely more pronounced in relatively small datasets.

In this section, we further investigate the impact of pretraining MAEs on high-quality, low-noise datasets. Specifically, we aim to find informative audio segments from recordings to retain only those segments containing at least one vocalizing species.
We mainly explore two filtering strategies: 1) \textbf{classification score filter}: thresholding the output scores from a pretrained classifier, and 2) \textbf{reconstruction filter}: thresholding the reconstruction loss from a pretrained MAE, which also can be viewed as hard sample mining.

\customparagraph{Classification Score Filter.} 
Since we train a single-label objective, when an audio segment is empty, a classifier should predict very low probability scores across species. We utilize a ViT-B model trained on the  full iNatSounds training set as an audio segment selector, determining whether to keep a segment based on the confidence score of its ground truth class. The underlying assumption is that segments with higher confidence are more informative. As shown in Figure~\ref{fig:classification_scores}, applying a confidence threshold of 0.88 retains approximately 50\% of the audio segments, corresponding to an average of 55\% of segments per recording in the iNatSounds dataset. To validate the quality of the filtered data, we train a MobileNetV3 model and observe that, despite using only 25\% of the original training data, the validation accuracy drops by just 1.5\%, remaining at 53.4\%. This suggests that the filtering strategy successfully preserves the informative patterns.
Next, we further pretrain masked autoencoders with the selected audio segments and finetune the models with either filtered data or full data. The results in Figure~\ref{fig:classification_scores} right show that finetuning with full data on iNatSounds achieves the highest top-1 accuracy of 59.5\%. For the models pretrained on the curated audio segments, finetuning on the selected audio segments again always yield higher accuracy than finetuning on the full dataset.

\customparagraph{Reconstruction Loss Filter.} 
Next, we explore a category agnostic filtering strategy. Empty audio segments may contain ambient sounds, but should have a relatively uniform or simpler structure than a segment with species sounds. Thus, such empty audio segments should be easier to reconstruct.
Specifically, we repurpose the ViT encoder trained during masked reconstruction as a selector of informative segments. We retain audio segments whose spectrograms have a high patch reconstruction loss, using difference thresholds as shown in Figure~\ref{fig:reconstruction_loss}. 
To evaluate this selection strategy, we train a MobileNetV3~\cite{howard2019searching} on the selected subsets. The results indicate a consistent drop in validation accuracy as the number of audio segments decreases. With only 10\% of the original training set, accuracy drops to 44.9\%, highlighting a significant performance degradation. This suggests that filtering based solely on reconstruction loss does not effectively preserve the most informative audio segments.
The pretraining and finetuning experiments also share a similar conclusion.
The results show that reconstruction loss filter does not work well on the data sample selection, and that the validation performance is still dominated by the decrease in the number of training audio segments.

\customparagraph{Takeaways from Data Curation.}
Our data curation strategies seemed to work qualitatively, and the relatively small drops in performance even after significant filtering of training data (Fig~\ref{fig:classification_scores} and Fig~\ref{fig:reconstruction_loss}, middle) suggests that we are efficiently picking the most informative segments. The lack of improvements with pretraining suggests that noisy, but more data is more beneficial for pretraining than curated, but less. This may actually be good news for the bioacoustics community, since it suggests potential benefits from scaling up to even noisier data, such as non-research-grade observations on iNaturalist.

\section{Conclusions and Future Work}
\label{sec:conclusion}

We study self-supervised MAE pretraining for fine-grained bioacoustic classification on the iNatSounds dataset.
Our results show that MAE models pretrained on general audio recordings, and even image datasets such as ImageNet, remain useful for the bioacoustic domain.
Continual self-supervised pretraining on a smaller domain-specific audio dataset further improves classification performance. However, additional pretraining within the same domain yields only marginal gains. 
We further investigate why in-domain pretraining provides limited benefits in our setting, in contrast to the success reported in some prior works.
For different post-training strategies such as linear probing and partial finetuning, we observe similar marginal benefits. Joint pretraining on mixed general and in-domain audio performs better, but the gains are again modest. 
Curating the data to reduce the influence of noisy or empty audio segments provides little benefit as well.

We conclude that supervised finetuning from off-the-shelf general domain pretrained models is the dominant method for moderate-sized datasets, and the actual benefit of in-domain pretraining appears with much larger scale data. 
In the future, we would like to study more advanced transfer learning methods for bioacoustic tasks, including parameter efficient finetuning (PEFT), and prototypical learning.
We hope our findings and lessons are useful for future explorations into MAEs for bioacoustics and help researchers make design choices based on their intended applications.
{
    \small
    \bibliographystyle{ieeenat_fullname}
    \bibliography{main}
}

% WARNING: do not forget to delete the supplementary pages from your submission 
% \input{sec/X_suppl}

\end{document}